\documentclass[reprint,superscriptaddress,nobibnotes,amsmath,amssymb,notitlepage,twocolumn,prl]{revtex4-1}

\usepackage{pdfpages}
\usepackage{etoolbox} 

\usepackage{graphicx}
\usepackage{dcolumn} 
\usepackage{bm} 
\usepackage{multirow}
\usepackage{mathptmx}
\usepackage{siunitx}
\usepackage{color}
\usepackage{hhline} 

\usepackage{enumitem}  

\usepackage[caption=false]{subfig}

\usepackage{hyperref}
\hypersetup{colorlinks=true, linkcolor=blue, citecolor=blue, urlcolor=blue}

\newcommand{\beq}[1]{\begin{equation}\label{#1}}
\newcommand{\eep}{\;.\end{equation}}
\newcommand{\eec}{\;,\end{equation}}
\newcommand{\eeq}{\end{equation}}

\DeclareMathOperator{\tr}{tr}

\newcommand{\lb}{\left(}
\newcommand{\rb}{\right)}

\newcommand*\tss[1]{\textsuperscript{#1}} 

\renewcommand{\a}{\alpha}
\renewcommand{\b}{\beta}
\newcommand{\g}{\gamma}
\renewcommand{\d}{\delta}
\newcommand{\ep}{\epsilon}

\newcommand{\x}{\chi}

\newcommand{\D}{\Delta}

\newcommand{\Om}{\Omega}

\DeclareMathAlphabet{\mathcal}{OMS}{cmsy}{m}{n} 

\newcommand{\Ef}{\mathcal{E}}   
\newcommand{\F}{\mathcal{F}}    
\newcommand{\bigO}{\mathcal{O}} 

\makeatletter
\patchcmd{\@outputpage@head}{\@ifx{\LS@rot\@undefined}{}{\LS@rot}}{}{}{}
\makeatother

\newcommand{\eo}{{\epsilon}_0}

\newcommand{\nt}{\tilde{\eta}}

\begin{document}


\title{Generalized relation between electromechanical responses at fixed voltage and fixed electric field}

\author{Daniel Bennett}
\email{dbennett@uliege.be}
\affiliation{Physique Th{\'e}orique des Mat{\'e}riaux, QMAT, CESAM, University of Li{\`e}ge, B-4000 Sart-Tilman, Belgium}

\author{Daniel Tanner}
\affiliation{Physique Th{\'e}orique des Mat{\'e}riaux, QMAT, CESAM, University of Li{\`e}ge, B-4000 Sart-Tilman, Belgium}
\affiliation{Universit{\'e} Paris-Saclay, CNRS, CentraleSup{\'e}lec, Laboratoire SPMS, 91190 Gif-sur-Yvette, France}

\author{Philippe Ghosez}
\affiliation{Physique Th{\'e}orique des Mat{\'e}riaux, QMAT, CESAM, University of Li{\`e}ge, B-4000 Sart-Tilman, Belgium}

\author{Pierre-Eymeric Janolin}
\affiliation{Universit{\'e} Paris-Saclay, CNRS, CentraleSup{\'e}lec, Laboratoire SPMS, 91190 Gif-sur-Yvette, France}

\author{Eric Bousquet}
\affiliation{Physique Th{\'e}orique des Mat{\'e}riaux, QMAT, CESAM, University of Li{\`e}ge, B-4000 Sart-Tilman, Belgium}

\begin{abstract}
We present a general relation between the electromechanical couplings of infinitesimal strain and electric field to arbitrary order, measured at fixed voltage and at fixed electric field. We show that the improper response at fixed field can be written as the strain derivative of the $n$\tss{th} order susceptibility tensor, and the proper response at fixed voltage drop can be written as the response at fixed field plus corrections for dilations and 90$^{\circ}$ rotations induced by strain. Our theory correctly reproduces the proper piezoelectric response and we go beyond with the electrostrictive response. We present first-principles calculations of the improper electrostrictive response at fixed field, and illustrate how the correction is used to obtain the proper response at fixed voltage. This distinction is of high importance given the recent interest in giant electrostrictors exhibiting electromechanical responses as large as the piezoelectric ones.
\end{abstract}

\maketitle

\section{Introduction}

The calculation of the piezoelectric response from first-principles density functional theory (DFT) simulations is nowadays standard, either using finite differences \cite{Vanderbilt2000} or density functional perturbation theory (DFPT) \cite{Wu2005}. 
However, care should be taken, as two possible piezoelectric responses can be obtained: the so-called “proper” and “improper” responses \cite{martin1972piezoelectricity,nelson1976linear,Vanderbilt2000,Wu2005}. 
The improper response is computed from DFT when calculations are performed at fixed electric field, whereas the proper response is typically measured experimentally, where the voltage and not the electric field is held fixed \cite{Vanderbilt2000,Stengel2009}. The distinction between fixed field and fixed voltage drop becomes necessary when strains are introduced.

\begin{figure}[ht!]
\centering
\hspace*{-1cm}
\includegraphics[width=\columnwidth]{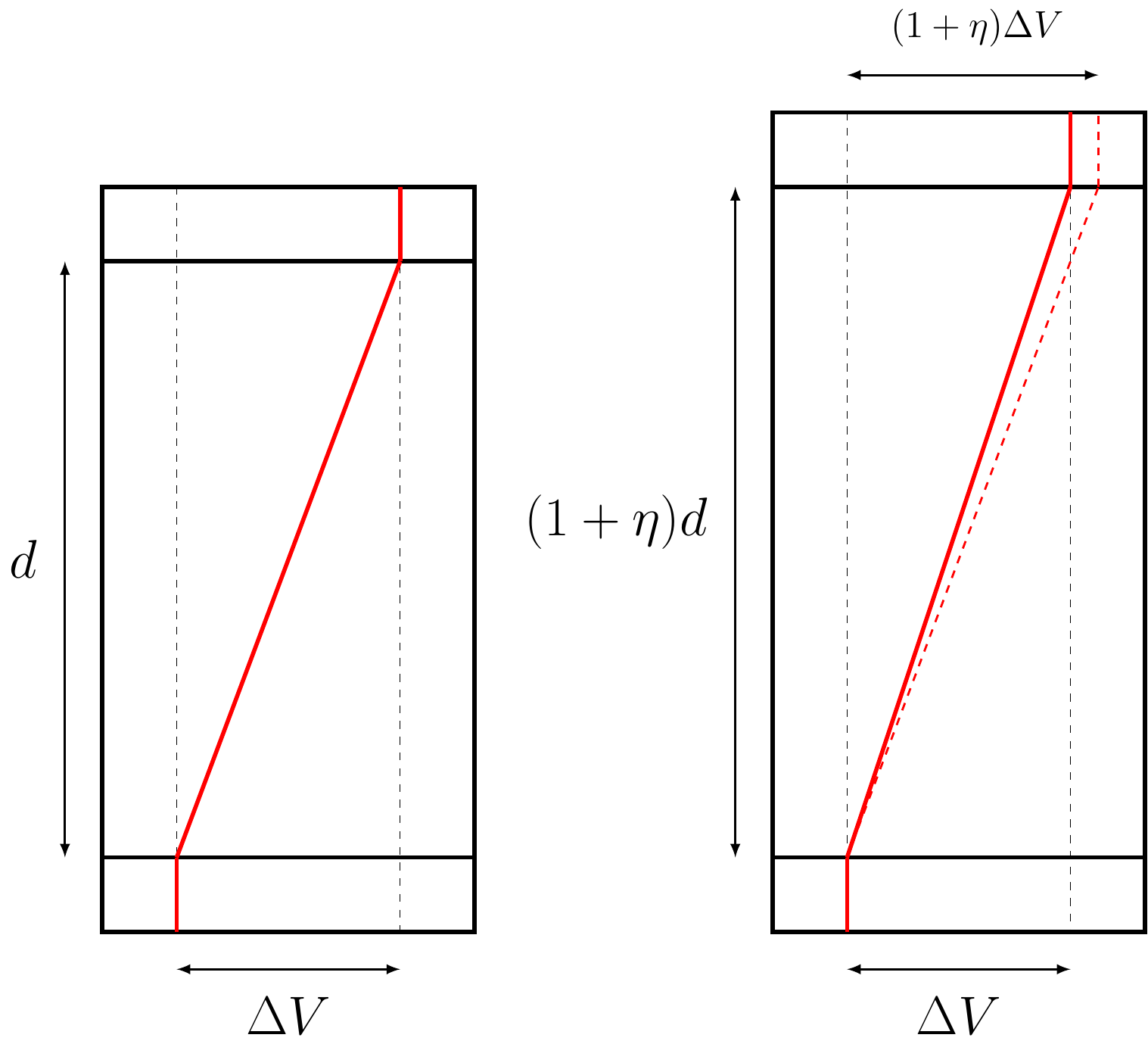}
\caption{Sketch of a dielectric material of thickness $d$ sandwiched between two metal plates. A potential difference $\D V$ is applied, resulting in an electric field of $\Ef = \frac{\D V}{d}$ inside the material. When a strain of $\eta$ is applied to the material, the field changes to $\Ef = \frac{\D V}{(1+\eta)d}$ if the potential difference is held fixed (experiment), or the potential difference changes to $(1+\eta)\D V$ if the field is held fixed (first-principles).}
\label{fig:intro-1}
\end{figure}

This problem is sketched in Fig.~\ref{fig:intro-1}: we have a dielectric material of thickness $d$ sandwiched between two metal plates, across which a potential drop $\D V$ is applied. The voltage drop is related to the electric field in the material via
\beq{eq:voltage_zero}
\D V = \Ef d
\eep
If a small strain $\eta$ appears in the material, changing the thickness to $(1+\eta)d$, Eq.~\eqref{eq:voltage_zero} becomes:
\beq{eq:voltage_eta}
\D V(\eta) = \Ef(1+\eta)d
\eep
If the field $\Ef$ is held fixed, which is typically the case in first-principles calculations, the potential difference has to change 
by $\eta \D V$. 
Thus when strain is introduced we need to compensate for this effect in order to hold the potential difference fixed. If the potential drop $\D V$ is fixed, then it is the reduced field $\Ef' = (1+\eta)\Ef$ rather than $\Ef$ which is held fixed in Eq.~\ref{eq:voltage_eta}. The voltage drop in each calculation will then be
\beq{}
\D V = (1+\eta)^{-1}\Ef' \cdot (1+\eta) d = \Ef' d
\eeq
which is independent of $\eta$.

\begin{figure}[ht!]
\centering
\hspace*{-1cm}
\includegraphics[width=\columnwidth]{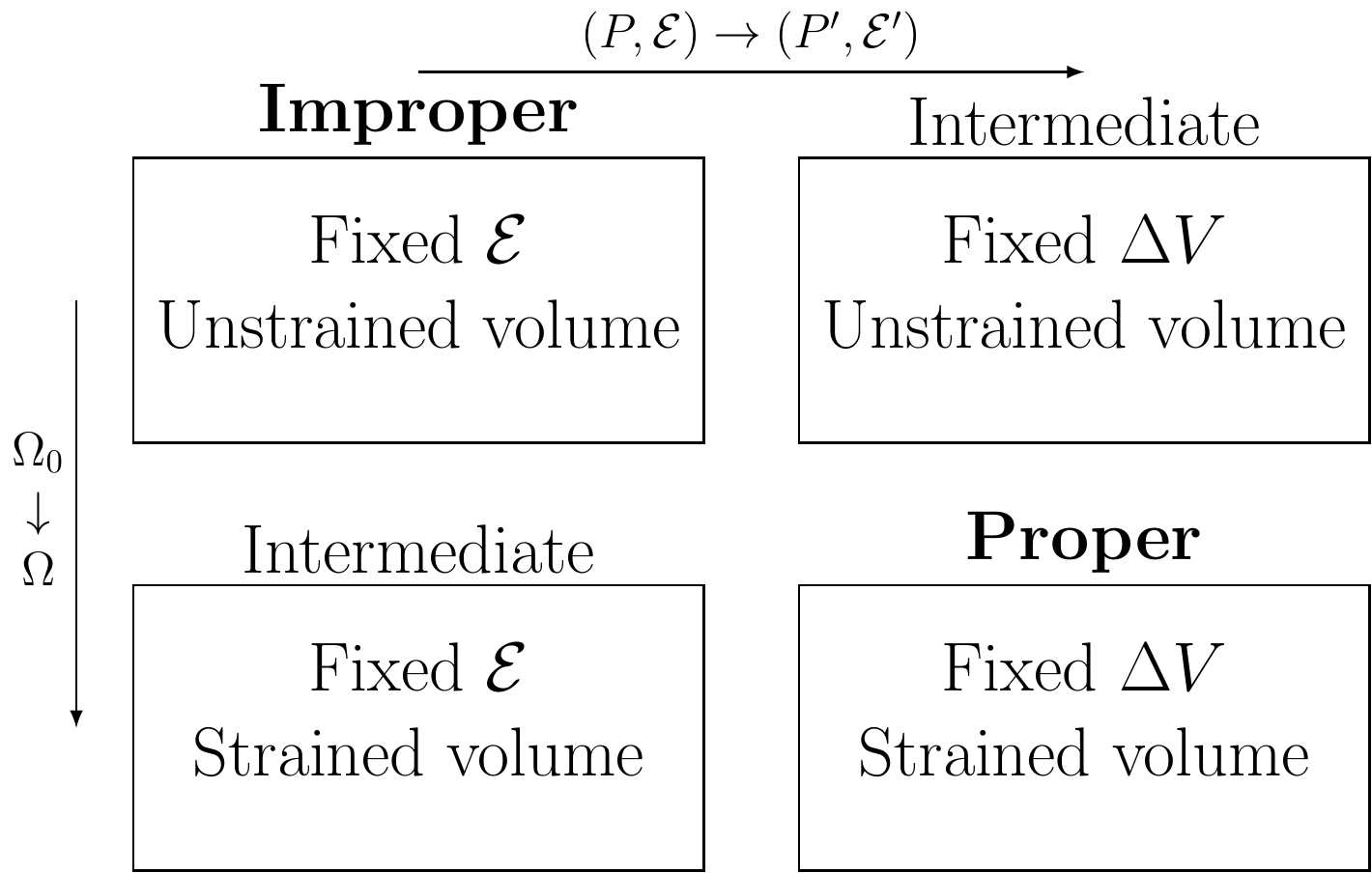}
\caption{Illustration of the four possible ways to define electromechanical responses, depending on the quantity held fixed ($\Ef$ or $\D V$) and the units of volume used ($\Om_0$ or $\Om$). The improper response is obtained by fixing $\Ef$ and working in units of unstrained volume, $\Om_0$, and the improper response is obtained by fixing $\D V$ and working in units of strained volume, $\Om$. Two additional `intermediate' responses which are neither proper nor improper are possible by switching the quantity held fixed or the units of volume.}
\label{fig:intro-2}
\end{figure}

In addition to the strain dependence of the relationship between $\Ef$ and $\D V$ (Eq.~\ref{eq:voltage_eta}), changes in the volume induced by strain can lead to different ways of measuring the energy per unit volume. Depending on whether calculations are done at fixed $\Ef$ or $\D V$, and whether the energy per unit volume is measured with respect to the equilibrium reference volume ($\Omega_0$) or the perturbed volume ($\Omega$), there are four possible ways to define an electromechanical response, see Fig.~\ref{fig:intro-2}: (i) the improper response, measured at fixed $\Ef$ and equilibrium volume $\Omega_0$, (ii) the proper response, measured at fixed $\D V$ and strained volume $\Omega$, (iii) an intermediate situation with fixed $\Ef$ and $\Omega$, and (iv) a second intermediate situation with fixed $\D V$ and $\Omega_0$ are used. This means that great care should be taken when comparing first-principles calculations of any electromechanical response to experimental measurements.

Although this problem is well-known in the context of piezoelectricity, its generalization to other electromechanical responses measured at fixed voltage and fixed electric field has not been reported. 
This generalization is becoming relevant, with increasing interest in higher-order couplings such as the electro-optic response and electrostriction.
Indeed, giant electrostrictors have recently been shown to give rise to electromechanical responses as large as piezoelectric ones \cite{yu2022defining,tanner2021divergent}, or even much larger: up to $2\times 10^{5} \ \si{pC/N}$ \cite{park2022induced}. 
However, no strict definition nor derivation of the proper versus improper responses have been given for electrostriction. In this paper, we derive an analogous relation for electrostriction and more generally any electromechanical response to an infinitesimal strain.

This paper is organized as follows: first, we review the relation between piezoelectric tensors measured at fixed electric field and fixed potential difference. 
The same methodology is then applied to obtain a corresponding relation for the electrostrictive tensor, and a general electromechanical coupling between an infinitesimal strain and electric field to arbitrary order. Detailed derivations are available in the Supplementary Material. We then use these relations to correct recent calculations of the electrostrictive response of rocksalt crystals in the literature \cite{tanner2020optimized}, focusing on MgO, and re-examine the comparison with experimental measurements.

\section{Theory}

For a crystal with mechanical degrees of freedom such as phonon displacements $u$ and macroscopic strain deformations $\nt$, which is subject to an applied electric field $\Ef$ \cite{nunes2001berry,souza2002first,umari2002ab}, the properties close to the equilibrium state can be described by Toupin's electric enthalpy density \cite{toupin1956elastic,ymeri1997toupin}:
\beq{eq:main_F_tot_unscaled}
\F_{\Om_0} = \F_{\text{cell}} - \lb \Ef\cdot P +\frac{1}{2}\eo \Ef^2\rb
\eec
where $\Om_0$ is the volume of the reference state ($\nt = 0$), $\F_{\text{cell}}$ is the zero-field Kohn-Sham energy per unit volume of the system, and $P$ is the total polarization of the system \cite{king1993theory,vanderbilt1993electric}. 
An homogeneous deformation results in a change in the positions $\mathbf{r}$ in the reference state: $\D r_\a = \nt_{\a\b}r_{\b}$. The symmetric part of $\nt$ is simply the strain tensor, $\eta$. In Ref.~\cite{Wu2005} it was shown that the proper piezoelectric tensor is symmetric under the exchange of the strain indices. Thus, although we consider general deformations $\nt$ in this paper, we use the terms strain and deformation interchangeably.

The various clamped responses ($u=0$) are defined as the derivatives of Eq.~\eqref{eq:main_F_tot_unscaled} about the equilibrium state: $\{\nt,\Ef\}=0$. 
For example, the piezoelectric tensor $e$ is defined as \cite{Wu2005}:
\beq{eq:main_e_improper}
e^{\text{(i)}}_{\a\b\g} \equiv - \left.\frac{\partial^2 \F_{\Om_0}}{\partial\Ef_{\a}\nt_{\b\g}}\right|_{\{\nt,\Ef\}=0}= \frac{\partial P_{\a}}{\partial\nt_{\b\g}}
\eec
where the superscript (i) denotes that this is the improper tensor, measured at fixed electric field. 
In order to obtain the proper tensor, at a fixed potential difference, two corrections must be made to Eq.~\eqref{eq:main_F_tot_unscaled}. 
First, the electrostatic energy must be scaled by $\frac{\Om}{\Om_0}$, where $\Om = \Om_0\det(I+\nt)$ is the volume after a strain deformation $\nt$. 
Secondly, in order to hold the potential difference fixed, we make the following change of variables to ``reduced coordinates'' for the electric field, polarization and displacement field $D$ \cite{stengel2013microscopic}:
\beq{eq:main_reduced}
\begin{split}
\Ef' &= \lb I+\nt\rb\Ef \\
P' &= \frac{\Om}{\Om_0}\lb I+\nt\rb^{-1} P \\
D' & = \frac{\Om}{\Om_0}\lb I+\nt\rb^{-1} D
\end{split}
\eec
which are equivalent to some of the reduced fields defined in Ref.~\cite{Stengel2009}, but here the units of each field are preserved \footnote{$\Ef'$ is equivalent to $\bar{\epsilon} = \mathbf{a}^T \Ef'$ but preserving the units of electric field, where the columns of $\mathbf{a}$ are the lattice vectors (including strain). $P'$ is equivalent to $p=\Omega \mathbf{b}^T P$, where the columns of $\mathbf{b}$ are the reciprocal lattice vectors, and similarly for $D'$. }. Thus, the enthalpy Eq.~\eqref{eq:main_F_tot_unscaled} becomes, in reduced coordinates:
\beq{eq:main_F_tot_reduced}
\F'_{\Om} =  \F_{\text{cell}} - \Ef'\cdot P' + \frac{\Om}{\Om_0}\frac{1}{2}\eo\Ef'^Tg^{-1}\Ef'
\eep
where $g=(I+\nt)^T(I+\nt)$. By differentiating we can see that $D'_{\a} = -\frac{\partial \F}{\partial\Ef'_{\a}}$, where we used $D = \eo\Ef + P$, although note that $D'\neq \eo\Ef'+P'$. 

The proper piezoelectric tensor, measured at fixed voltage drop, is defined as:
\beq{eq:main_e_proper}
e_{\a\b\g} \equiv -\left.\frac{\partial^2 \F'_{\Om} }{\partial \Ef'_{\a} \partial \nt_{\b\g}}\right|_{\{\nt,\Ef'\}=0} = \frac{\partial P'_{\a}}{\partial\nt_{\b\g}}
\eep
The term linear in $\Ef$ is absent in Eq.~\eqref{eq:main_e_proper} as the derivation is carried out about the equilibrium state ($\nt,\Ef\to 0$).
Summation convention is assumed, using Latin letters for dummy indices and Greek letters for free indices. 
Next, we explicitly evaluate the derivative in order to write Eq.~\eqref{eq:main_e_proper} in terms of Eq.~\eqref{eq:main_e_improper}:
\beq{}
\begin{split}
e_{\a\b\g} &= \frac{\partial \left[  \det\lb I + \nt\rb \lb I + \nt \rb^{-1}_{\a j} P_{j} \right] }{\partial\nt_{\b\g}} \\
 &= \frac{\partial \left[  \lb 1 + \nt_{ii}\rb \lb \d_{\a j} - \nt_{\a j}\rb P_{j} + \bigO(\nt^2)\right] }{\partial\nt_{\b\g}} \\
 &= \frac{\partial P_{\a}}{\partial\nt_{\b\g}} + P_{\a}\d_{\b\g} - P_{\g}\d_{\a\b} + \nt_{ii}\frac{\partial P_{\a}}{\partial\nt_{\b\g}} - \nt_{\a j}\frac{\partial P_{j}}{\partial\nt_{\b\g}}
\end{split}
\eep
Going from the first line to the second line, we used the expansion $\lb I + \nt \rb^{-1} = I - \nt + \bigO(\nt^2)$, truncating to linear order, assuming the strain is infinitesimal. For the same reason, we use $\det(I+\nt) = 1+\tr{(\nt)} + \bigO(\nt^2)$ to simplify the volume term. Note that it is important to do this \textit{before} differentiating, otherwise contributions which are quadratic in strain will be retained. After differentiating, we set $\Ef,\nt\to 0$, because the derivatives are defined about the equilibrium state with zero strain and applied field. The relation between piezoelectric tensors measured at fixed voltage and at fixed field is then:
\beq{eq:main_e_proper_improper}
e_{\a\b\g} = e^{\text{(i)}}_{\a\b\g} + P_{\a}\d_{\b\g} - P_{\g}\d_{\a\b}
\eep
As discussed in Ref.~\onlinecite{Vanderbilt2000}, the first term is a correction for dilations of the polarization induced by strain. The second term is a correction for rotations of the polarization by 90$^{\circ}$, i.e.~permutations of the indices. Additionally, a problem unique to the piezoelectric tensor is that Eq.~\eqref{eq:main_e_improper} is sensitive to the branch on which the polarization is measured (quantum of polarization), but this is remedied by using Eq.~\eqref{eq:main_e_proper_improper}, which is branch invariant \cite{Vanderbilt2000}. 

\subsection{Electrostriction}

The improper strain electrostrictive tensor $m$, measured at fixed electric field, is defined as follows: 
\beq{eq:main_m_improper}
m^{(\text{i})}_{\a\b\g\d} \equiv -\frac{1}{2}\left.\frac{\partial^3 \F_{\Om_0}}{\partial \Ef_{\a} \partial \Ef_{\b} \partial \nt_{\g\d}}\right|_{\{\nt,\Ef\}=0} = \frac{\partial \ep^{(\text{i})}_{\a\b}}{\partial \nt_{\g\d}} 
\eec
where $\ep^{(\text{i})}_{\a\b} \equiv \frac{\partial D_{\a}}{\partial \Ef_{\b}}$ is the improper permittivity tensor, measured at fixed electric field calculated about the equilibrium state. 
The proper electrostrictive tensor, measured at fixed voltage, is defined as follows:
\beq{eq:main_m_proper}
m_{\a\b\g\d} \equiv -\frac{1}{2}\left.\frac{\partial^3 \F'_{\Om}}{\partial \Ef'_{\a} \partial \Ef'_{\b} \partial \nt_{\g\d}}\right|_{\{\nt,\Ef'\}=0} = \frac{\partial \ep_{\a\b}}{\partial\nt_{\g\d}}
\eec
where $\ep_{\a\b} \equiv \frac{\partial D'_{\a}}{\partial \Ef'_{\b}}$ is the proper permittivity tensor, measured at fixed voltage. Immediately we see that Eqs.~\eqref{eq:main_m_improper} and \eqref{eq:main_m_proper} are analogous to Eqs.~\eqref{eq:main_e_improper} and \eqref{eq:main_e_proper}, but with polarization replaced by dielectric permittivity. As before, we must write Eq.~\eqref{eq:main_m_proper} in terms of Eq.~\eqref{eq:main_m_improper}. In order to do this, we must first relate the proper and improper permittivity tensors:
\beq{eq:main_epsilon_proper_improper}
\ep_{\a\b} = (1+\nt_{ii})\ep^{(\text{i})}_{\a\b} - \ep^{(\text{i})}_{\a i}\nt_{i\b} - \nt_{\a j}\ep^{(\text{i})}_{j\b} + \bigO(\nt^2)
\eep
Note that $\ep_{\a\b} = \ep^{(\text{i})}_{\a\b}$ at $\nt=0$, which is expected. However, it is clear that their derivatives with respect to strain will not be equal at $\nt=0$. Inserting Eq.~\eqref{eq:main_epsilon_proper_improper} into Eq.~\eqref{eq:main_m_proper}:
\beq{}
\begin{split}
m_{\a\b\g\d} & = \frac{\partial \ep_{\a\b}}{\partial\nt_{\g\d}} \\
 & = \frac{\partial }{\partial\nt_{\g\d}}\left[ (1+\nt_{ii})\ep^{(\text{i})}_{\a\b} - \ep^{(\text{i})}_{\a i}\nt_{i\b} - \nt_{\a j}\ep^{(\text{i})}_{j\b} + \bigO(\nt^2) \right] \\
  & = \frac{\partial \ep^{(\text{i})}_{\a\b}}{\partial\nt_{\g\d}} + \ep^{(\text{i})}_{\a\b} \d_{\g\d} - \ep^{(\text{i})}_{\a \g}\d_{\b\d} - \ep^{(\text{i})}_{\d\b} \d_{\g\a} \\
\end{split}
\eep
Hence, we obtain an analogous relation between the proper and the improper response to Eq.~\eqref{eq:main_e_proper_improper} for the electrostriction:
\beq{eq:main_m_proper_improper}
m_{\a\b\g\d} = m^{(\text{i})}_{\a\b\g\d} + \ep_{\a\b}\d_{\g\d} - \ep_{\a\g}\d_{\b\d} - \ep_{\d\b}\d_{\g\a}
\eec
where we have dropped the (i) superscript on the permittivities because we have set $\nt\to 0$ and hence $\ep^{(\text{i})} = \ep$. The corrections are similar to those appearing in the proper-improper relation for the piezoelectric tensors: the first term is a correction for changes in the permittivity induced by strain. The second term is a correction for permutations of the indices. Note that there are two of these terms, since $\ep$ is a rank 2 tensor.

\subsection{Generalization to all orders of electromechanical coupling}

The proper-improper relation can be generalized to electromechanical couplings which are linear in strain and to any order in electric field. The (improper) coupling between linear strain and $n$\tss{th} order electric field is described by the following tensor:
\beq{eq:main_A_improper}
\begin{split}
A^{(\text{i})}_{\Ef_1\ldots \Ef_n\nt_1\nt_2} &\equiv -\frac{1}{n!}\left.\frac{\partial^{n+1}\F_{\Om_0}}{\partial\Ef_{\Ef_1}\ldots \partial\Ef_{\Ef_n}\partial\nt_{\nt_1\nt_2}}\right|_{\{\nt,\Ef\}=0} \\
&= \frac{\partial\x^{(\text{i})}_{\Ef_1\ldots\Ef_n}}{\partial\nt_{\nt_1\nt_2}}
\end{split}
\eec
measured at fixed field, where $n=1$ corresponds to the piezoelectric tensor and $n=2$ corresponds to the electrostrictive tensor. 
We can see that in general $A$ is the first strain derivative of the $n$\textsuperscript{th} order susceptibility tensor:

\beq{eq:main_chi_general_improper}
\x^{(\text{i})}_{\Ef_1\ldots\Ef_n} \equiv \left\{\begin{array}{cr}
        P_{\Ef_1}, & \quad n=1\\[10pt]
        \frac{\partial D_{\Ef_1}}{\partial\Ef_{\Ef_2}}, & \quad n=2\\[10pt]
        \frac{\partial^{n-1}P_{\Ef_1}}{\partial\Ef_{\Ef_2}\ldots \partial\Ef_{\Ef_n}}, & \quad n > 2
        \end{array}\right.
\eeq
defined to be the polarization and permittivity for ${n=1,2}$, respectively. The proper tensor, measured at fixed voltage, is defined as:
\beq{eq:main_A_proper}
\begin{split}
A_{\Ef_1\ldots \Ef_n\nt_1\nt_2} &\equiv -\frac{1}{n!}\left. \frac{\partial^{n}\F'_{\Om}}{\partial\Ef'_{\Ef_1}\ldots \partial\Ef'_{\Ef_n}\partial\nt_{\nt_1\nt_2}}\right|_{\{\nt,\Ef'\}=0} \\
&= \frac{\partial\x_{\Ef_1\ldots\Ef_n}}{\partial\nt_{\nt_1\nt_2}} \\
\end{split}
\eec
where
\beq{eq:main_chi_general_proper}
\x_{\Ef_1\ldots\Ef_n} \equiv \left\{\begin{array}{cr}
        P'_{\Ef_1}, & \quad n=1\\[10pt]
        \frac{\partial D'_{\Ef_1}}{\partial\Ef'_{\Ef_2}}, & \quad n=2\\[10pt]
        \frac{\partial^{n-1}P'_{\Ef_1}}{\partial\Ef'_{\Ef_2}\ldots \partial\Ef'_{\Ef_n}}, & \quad n > 2
        \end{array}\right.
\eep
Again, the aim is to write Eq.~\eqref{eq:main_A_proper} in terms of Eq.~\eqref{eq:main_A_improper}. 
In order to do this, we must first obtain the relation between the $n$\textsuperscript{th} order susceptibility tensors:
\beq{}
\begin{split}
\x_{\Ef_1\ldots\Ef_n} &= \frac{\partial^{n-1}P'_{\Ef_1}}{\partial\Ef'_{\Ef_2}\ldots \partial\Ef'_{\Ef_n}} \\
& = \left[\prod_{i=2}^n \lb I+\nt\rb^{-1}_{k_i\Ef_i}\right]\frac{\partial^{n-1}P'_{\Ef_1}}{\partial\Ef_{k_2}\ldots \partial\Ef_{k_n}}\\
& = \left[\prod_{i=1}^n \lb I+\nt\rb^{-1}_{k_i\Ef_i}\right] \det(I+\nt)\x^{(\text{i})}_{k_1\ldots k_n}
\end{split}
\eeq
Expanding $\det(I+\nt)$ and the products of $(I+\nt)^{-1}$, and truncating to linear order, we get the following expression:
\beq{eq:main_chi_general_proper_improper}
\x_{\Ef_1\ldots\Ef_n} = (1+\nt_{ii})\x^{(\text{i})}_{\Ef_1\ldots\Ef_n} - \sum_{i=1}^{n} \x^{(\text{i})}_{\Ef_1\ldots k_i \ldots\Ef_n}\nt_{k_i\Ef_i}
\eep
Eq.~\eqref{eq:main_chi_general_proper_improper} is the relation between $n$\textsuperscript{th} order susceptibility tensors in reduced and unreduced coordinates at finite strain. 
It is reassuring to see that when $\nt\to 0$ we have $\x_{\Ef_1\ldots\Ef_n} = \x^{(\text{i})}_{\Ef_1\ldots\Ef_n}$, as expected. Obtaining the proper-improper relation for $A$ is now straightforward:
\beq{eq:main_A_proper_improper}\resizebox{\columnwidth}{!}{$
\begin{split}
A_{\Ef_1\ldots \Ef_n\nt_1\nt_2} &= \frac{\partial}{\partial \nt_{\nt_1\nt_2}} \lb (1+\nt_{ii})\x^{(\text{i})}_{\Ef_1\ldots\Ef_n} - \sum_{i=1}^{n} \x^{(\text{i})}_{\Ef_1\ldots k_i \ldots\Ef_n}\nt_{k_i\Ef_i}\rb\\
& = A^{(\text{i})}_{\Ef_1\ldots \Ef_n\nt_1\nt_2} + \d_{\nt_1\nt_2}\x_{\Ef_1\ldots\Ef_n} - \sum_{i=1}^{n} \x_{\Ef_1\ldots \nt_1 \ldots\Ef_n}\d_{\nt_2\Ef_i}
\end{split}
$}
\eeq
where $\nt_1$ is in the $i$\tss{th} position. There is one term for dilations induced by strain and $n$ terms corresponding to permutations of the $n$ indices of the susceptibility. For $n=1,2$, Eqs.~\eqref{eq:main_e_proper_improper} and \eqref{eq:main_m_proper_improper} are reproduced, respectively.

\subsection{First-principles calculations}

\begin{figure}[ht!]
\centering
\includegraphics[width=\columnwidth]{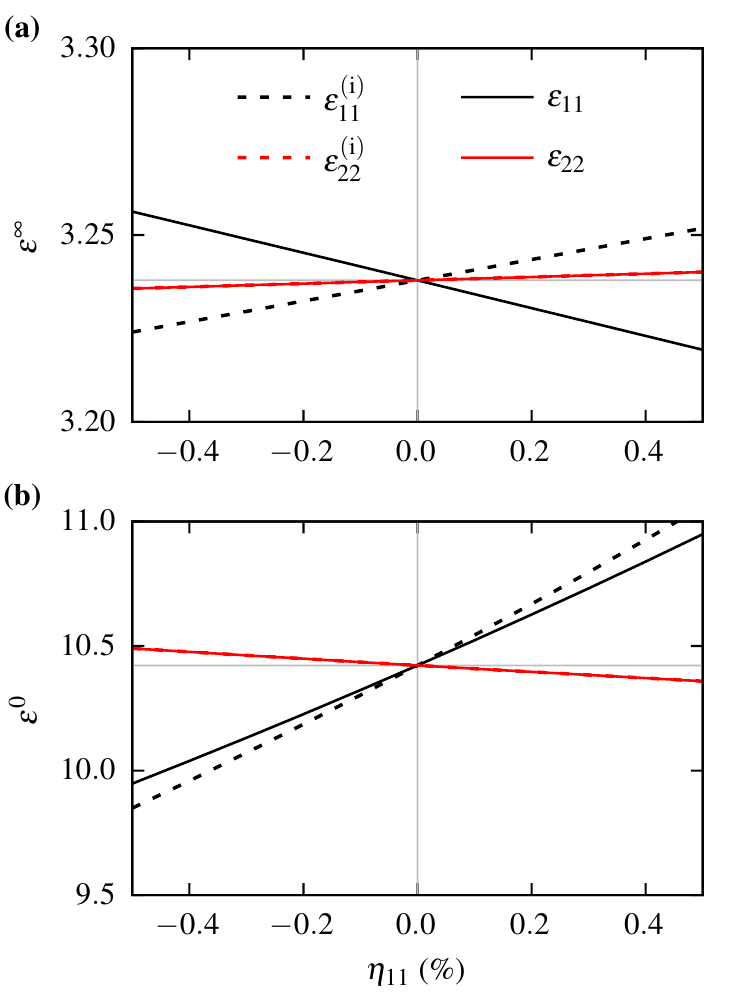}
\caption{Improper (relative) dielectric permittivity of MgO as a function of applied strain $\eta_{11}$ (dashed lines) \cite{tanner2020optimized}, and the (relative) proper permittivity obtained using Eq.~\eqref{eq:main_epsilon_proper_improper} (solid lines), for \textbf{(a)}: the electronic permittivity $\ep^{\infty}$ (clamped ions) and \textbf{(b)}: the relaxed-ion permittivity $\ep^0$.}
\label{fig:epsilon}
\end{figure}

\begin{table*}[ht!]
\renewcommand*{\arraystretch}{2}
\setlength{\tabcolsep}{10pt}
\begin{center}
\begin{tabular}{| c | c | c | c | c | c | c |}
\hline\hline
\multirow{2}{*}{Material} & \multirow{2}{*}{$M$ (\si{pm^2/V^2})} & Improper & Intermediate & Intermediate & Proper  & \multirow{2}{*}{Experiment} \\
 & & fixed $\Ef$ & $(\Ef,P)\to (\Ef',P')$ & $\Om_0 \to \Om$ & fixed $\D V$ & \\ \hline
\multirow{2}{*}{MgO} & $M_{11}$ & 1829 & 2460 & 1514 & 2144 & 2020$^{\text{a}}$ \\ \hhline{~|-|-|-|-|-|-|}
 & $M_{12}$ & -199 & -199 & -514 & -514 & - \\ \hline
 \multirow{2}{*}{LiCl} & $M_{11}$ & 12613 & 15836 & 11001 & 14224 & 46200$^{\text{b}}$ \\ \hhline{~|-|-|-|-|-|-|}
 & $M_{12}$ & -1281 & -1281 & -2892 & -2892 & -18200$^{\text{b}}$ \\ \hline
 \multirow{2}{*}{LiF} & $M_{11}$ & 4640 & 5941 & 3989 & 5290 & 5230$^{\text{c}}$ \\ \hhline{~|-|-|-|-|-|-|}
 & $M_{12}$ & -501 & -501 & -1151 & -1151 & -1730$^{\text{c}}$ \\ \hline
 \multirow{2}{*}{NaCl} & $M_{11}$ & 4572 & 6516 & 3600 & 5544 & 4030$^{\text{c}}$ \\ \hhline{~|-|-|-|-|-|-|}
 & $M_{12}$ & -483 & -483 & -1455 & -1455 & -1030$^{\text{c}}$ \\ \hline
\hline
\end{tabular}
\end{center}
\caption{Measurements of the improper, proper and intermediate electrostrictive responses $M_{11}\equiv M_{1111}$ and $M_{12}\equiv M_{2211}$ of MgO, LiCl, LiF and NaCl, compared to experimental measurements: a = Ref.~\cite{sundar1996interferometric}, b = Ref.~\cite{kucharczyk1987estimation} and c= Ref.~\cite{ScHa99}.}
\label{table:m}
\end{table*}

We illustrate the proper-improper relation for electrostriction using first-principles calculations of the electrostrictive response of MgO, LiCl, LiF and NaCl. Typically, it is the electrostrictive response to stress which is measured experimentally, because it can be measured at fixed voltage \cite{yu2022defining}. The electrostrictive response to a stress is defined as ${M_{\a\b\g\d} = \frac{1}{2}\frac{\partial \ep_{\a\b}}{\partial X_{\g\d}}}$, and is related to the electrostrictive response to strain used in the previous section by ${M_{\a\b\g\d} = s_{\a\b\mu\nu}m_{\mu\nu\g\d}}$, where $s$ is the compliance tensor, the inverse of the elastic tensor. In order to obtain the electrostrictive response to stress at fixed voltage, we first measure the electrostrictive response to strain $m$, apply the correction Eq.~\eqref{eq:main_m_proper_improper}, and then contract with the compliance tensor to obtain $M$.

Fig.~\ref{fig:epsilon} shows the permittivity of MgO versus strain as obtained with the {\sc abinit} code \cite{gonze2016,gonze2020}, following the methodology of Ref.~\cite{tanner2020optimized}. Norm conserving pseudopotentials from PseudoDojo were used \cite{Setten2018}, and the PBEsol functional was used to treat exchange-correlation interactions \cite{pbesol}. A cutoff energy of 50 Ha was used to truncate the plane-wave basis, and a $k$-point grid of $8\times 8 \times 8$ was used to sample the Brillouin zone. The dielectric permittivity was then calculated using DFPT. The electronic or clamped-ion permittivity $\ep^{\infty}$ was first obtained, and then phonon calculations were performed in order to obtain the relaxed-ion permittivity $\ep^0$. 

The dashed lines show the components of the permittivity tensor obtained directly from {\sc abinit}, which are in unreduced units and therefore improper. The components of the proper dielectric tensor, indicated by the solid lines, are obtained using Eq.~\eqref{eq:main_epsilon_proper_improper}. Note that for $\ep_{22}$, the proper and improper values are identical, which can be verified using Eq.~\eqref{eq:main_epsilon_proper_improper}. The improper electrostrictive response $m^{\text{(i)}}$ is obtained by taking the slope of $\ep^{\text{(i)}}$ about $\eta=0$, and the proper response $m$ is obtained by taking the slope of $\ep$ about $\eta=0$. The two intermediate values, obtained by either using reduced variables for the electric field or units of strained volume, were calculated using the corrections in Fig.~\ref{fig:mapping} in order to illustrate their individual contributions to the total correction. The electrostrictive response to stress $M$ was then obtained by contracting $m$ with the compliance tensor in each case. In Table \ref{table:m} we show a comparison of the electrostrictive responses for the proper, improper and intermediate cases, as well as experimental measurements.

\section{Discussion and Conclusions}

\begin{figure*}[ht!]
\centering
\hspace*{-1cm}
\includegraphics[width=\textwidth]{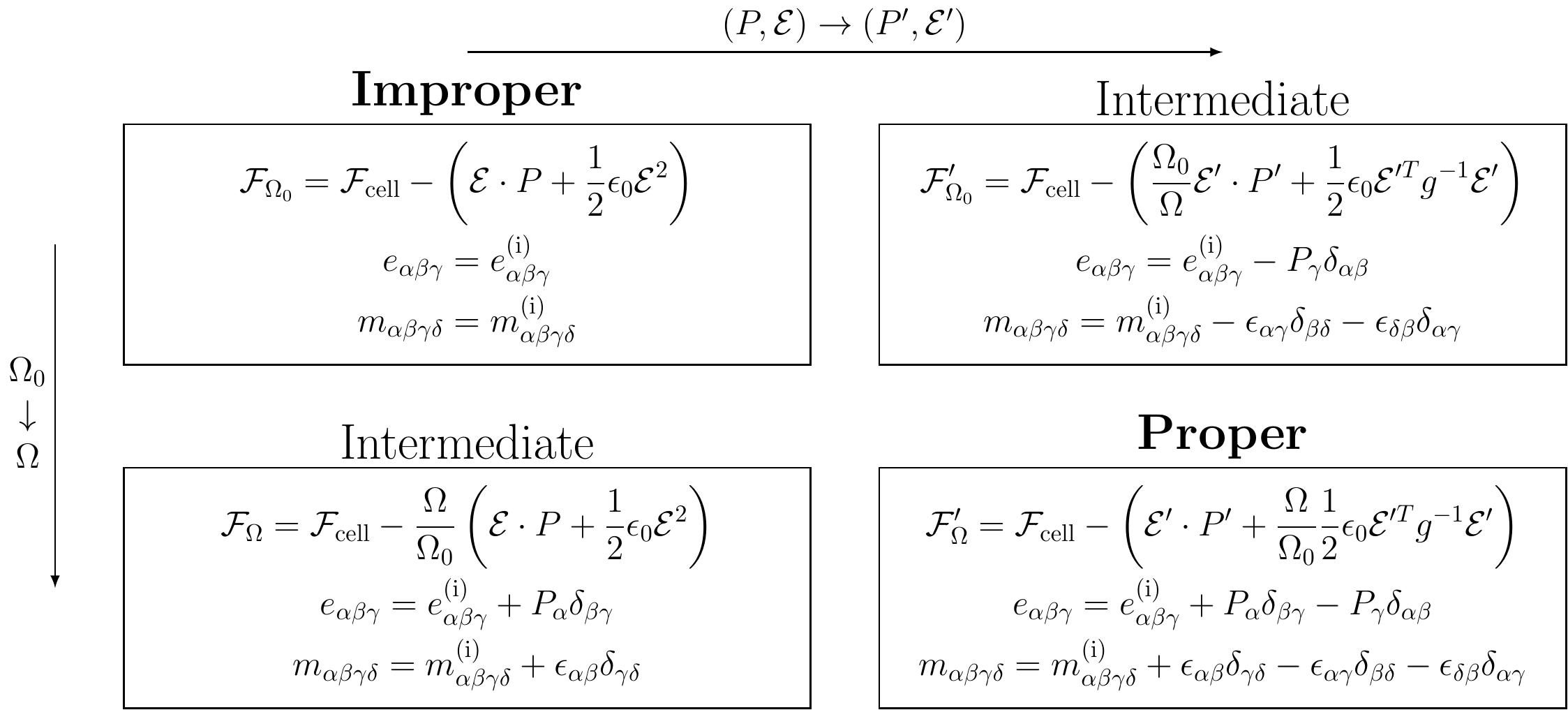}
\caption{Summary of the relations between the free energies, piezoelectric and electrostrictive responses for the four possible scenarios illustrated in Fig.~\ref{fig:intro-2}.}
\label{fig:mapping}
\end{figure*}

In this paper, we highlighted the difficulties in comparing electromechanical responses obtained from first-principles calculations to those obtained from experimental measurements. 
In the context of piezoelectricity, this problem is known as the `proper-improper' relation, which may not be intuitive as `proper' and `improper' are ambiguous terms. We believe it is more physically insightful to reformulate the problem in terms of responses measured at fixed electric field and fixed voltage drop. 
Similar to the well-known relation for piezoelectricity, we have derived the relations for electrostrictive and general electromechanical responses to infinitesimal strain, measured at fixed field and at fixed voltage drop.

Depending on whether the field or voltage is held fixed, and whether the energy is given per unstrained or strained unit volume, there are four different ways to define an electromechanical response: the so-called `improper` (fixed field) and `proper` (fixed voltage) responses, and two intermediate ones.
We summarize these possibilities in Fig.~\ref{fig:mapping} to help the reader in identifying each case. In order to correctly compare measurements from first-principles calculations with experimental measurements, it is important to identify which response from Fig.~\ref{fig:mapping} has been measured in each case and, if necessary, make the appropriate correction.

Because the electrostrictive response is proportional to a third derivative of the free energy (Eq.~\eqref{eq:main_m_improper}), it can be expressed in several different ways:
\beq{}
m_{\a\b\g\d} = \frac{\partial \ep_{\a\b}}{\partial \eta_{\g\d}} = \frac{\partial^2 X_{\g\d}}{\partial \Ef_{\a}\partial\Ef_{\b}} = \frac{\partial e_{\b\g\d}}{\partial \Ef_{\a}}
\eep
Therefore, the electrostrictive response can be calculated from first-principles using three different approaches:
\begin{enumerate}[label=(\roman*)]
\item by calculating the linear evolution of the dielectric permittivity with respect to stress or strain.
\item by calculating the quadratic stress induced by an applied electric field.
\item by calculating the linear of evolution of the piezoelectric tensor in response to an applied electric field.
\end{enumerate}
Second derivatives of the free energy can be calculated from first-principles calculations using second-order DFPT, and in {\sc abinit}, the proper piezoelectric response is automatically obtained because reduced units $(\Ef',P')$ are used for the electric field and polarization. However, third-order DFPT for the electrostrictive response is not currently implemented in any widely available DFT code, and therefore in all three of the above methods, at least one of the derivatives must be evaluated using finite difference methods. Thus, a correction must be applied in order to obtain the proper response at fixed voltage.

With method (i), first introduced in Ref.~\cite{tanner2020optimized} and used in this study, the electrostrictive response is obtained by measuring the derivative of the permittivity with respect to strain. The permittivity can easily be obtained from a single DFPT calculation, avoiding the need to perform a set of finite field calculations for each value of strain. Taking the derivative of the dielectric permittivity output from the code from finite differences will yield the improper response, i.e.~Eq.~\eqref{eq:main_m_improper} \cite{tanner2020optimized}. The proper response can be obtained by taking care to write the permittivity in terms of $\Ef'$ and $P'$, i.e.~using Eq.~\eqref{eq:main_epsilon_proper_improper}, $m \approx \frac{ \ep(\nt_+) - \ep(\nt_-)}{\nt_+ - \nt_-}$, or by applying the correction (Eq.~\eqref{eq:main_m_proper_improper}) to the improper response. We propose that method (i) is the most straightforward method for calculating the electrostrictive response from first-principles.

With method (ii), the electrostrictive response $M$ is obtained by measuring the quadratic strain response to an applied field, and measuring the curvature about zero field. This has previously been the most widely used method to measure electrostrictive responses from first-principles. However, if finite electric field calculations are performed \cite{kornev2010electrostriction,PeKa12,tanner2020optimized}, the improper response will be obtained and a correction will be needed. The proper response can be directly obtained by fixing the reduced electric field $\Ef'$ or the displacement field \cite{cancellieri2011electrostriction,JiZh16,tanner2020optimized}, both of which are possible in {\sc abinit}, but not in most other widely available DFT codes. Additionally, a geometry relaxation calculation must be performed for each field value in order to find the induced strain, making this method more computationally expensive than method (i).

With method (iii), the electrostrictive response could be obtained by measuring the change in the piezoelectric tensor in response to an applied field \cite{yimnirun2002electrostrictive}. To our knowledge, the electrostrictive response has not yet been calculated from first-principles in this way. However, if the proper piezoelectric response is obtained, and the reduced electric field $\Ef'$ or displacement field is held fixed, then the proper piezoelectric response should be obtained.

For the electronic permittivity, the sign of the slope changes when correcting from improper to proper. The reason for this is that the electronic permittivity is small and more sensitive to strain than the larger relaxed-ion permittivity (i.e. including phonon contributions), and the correction appears relatively larger.

In each case, the electrostrictive response is larger after correcting to the fixed voltage case. Agreement with experimental measurements is improved for the case of LiF, but not for MgO or NaCl. For the case of LiCl, the reported values in Ref.~\cite{ScHa99} are an order of magnitude larger than the typical values measured in other experimental studies. Experimental measurements of electrostrictive responses tend to vary significantly, in part due to the various measurement techniques. For example, the $M_{11}$ electrostrictive coefficient of MgO has been reported to be 2020 \si{pm^2/V^2} in Ref.~\cite{sundar1996interferometric} and about four times less (550 \si{pm^2/V^2}) in Ref.~\cite{yimnirun2002electrostrictive}. For SrF$_2$, the situation is even more dramatic as experimental reports disagree even on the \textit{sign} of the longitudinal $M$ coefficient: -1160 \si{pm^2/V^2} from Ref.~\cite{meng1984electrostriction} and +260 \si{pm^2/V^2} from Ref.~\cite{van1992measurement}. As a consequence, the agreement with experimental values has to be considered with care. One typical source of error between first-principles calculations and experimental measurements is that first-principles calculations are typically done at zero temperature, whereas experimental measurements are performed at larger temperatures (in this case room temperature \cite{sundar1996interferometric}).  However, the evolution of the electrostrictive coefficients with respect to temperature is normally very small, provided a phase transition does not occur \cite{tanner2021divergent}.

In spite of this, it is important to emphasise that the fixed field and fixed voltage responses are \textit{physically different quantities}. When comparing first-principles calculations with experimental measurements, it is of primary importance to first ensure that the correct physical quantities are being compared.

\section*{Acknowledgements}

The authors thank M. Stengel for helpful discussions. 
DB acknowledges funding from the University of Li{\'e}ge under special funds for research (IPD-STEMA fellowship programme). 
Computational resources have been provided by the Consortium des \'Equipements de Calcul Intensif (C\'ECI), funded by the Fonds de la Recherche Scientifique de Belgique (F.R.S.-FNRS) under Grant No.~2.5020.11 and by the Walloon Region and using the DECI resource BEM based in Poland at Wrocław with support from the PRACE OFFSPRING project.
EB acknowledges FNRS for support and DT aknowledge ULi{\'e}ge Euraxess support.
This work was also performed using HPC resources from the “M\'esocentre” computing centre of CentraleSup\'elec and \'Ecole Normale Sup\'erieure Paris-Saclay supported by CNRS and R\'egion \^Ile-de-France (http://mesocentre.centralesupelec.fr/).
Financial support is acknowledged from public grants overseen by the French National Research Agency (ANR) in the  ANR-20-CE08-0012-1 project and as part of the ASTRID program (ANR-19-AST-0024-02). 


%

\clearpage

\includepdf[pages={1}]{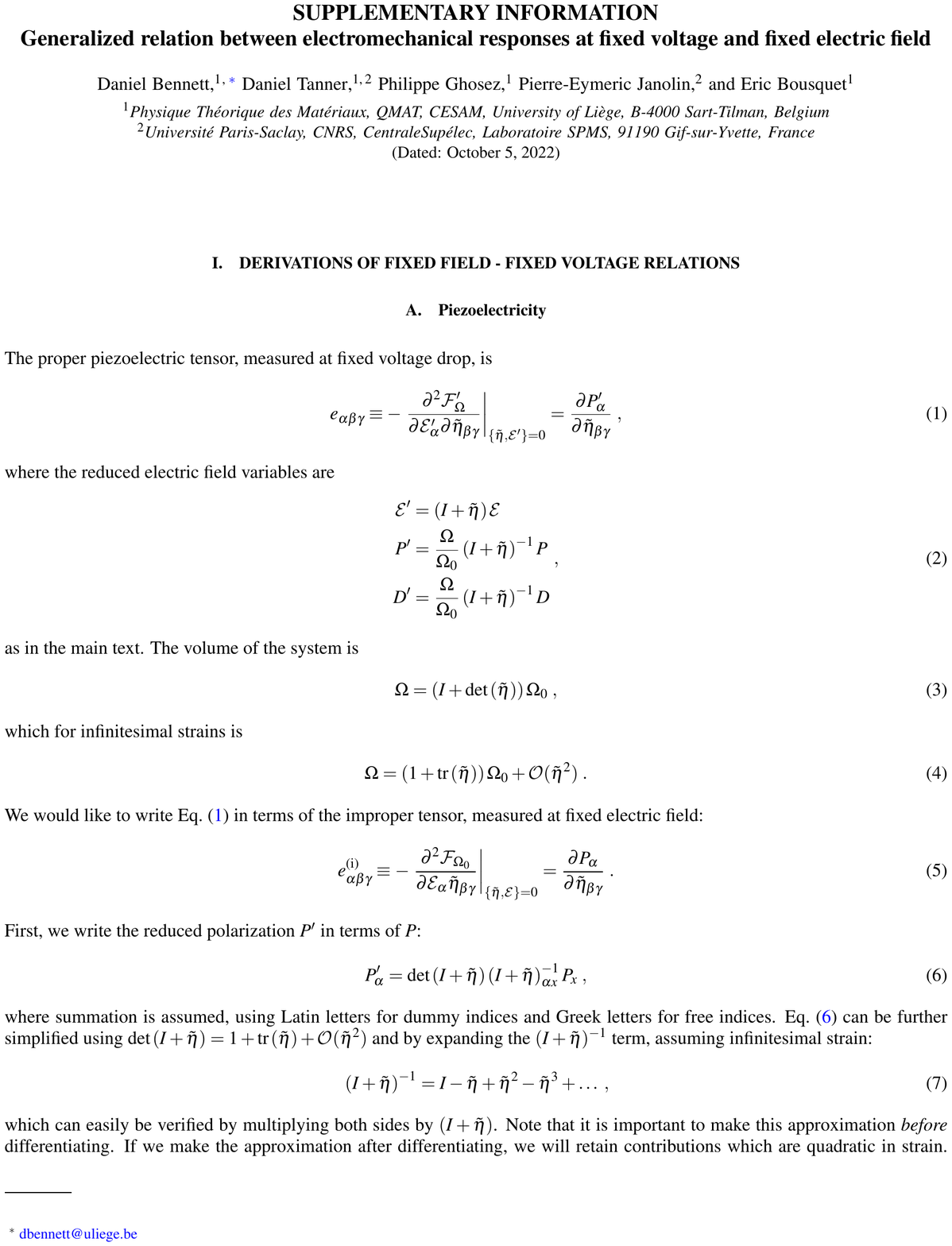}
\clearpage
\includepdf[pages={2}]{./SI.pdf}
\clearpage
\includepdf[pages={3}]{./SI.pdf}
\clearpage
\includepdf[pages={4}]{./SI.pdf}

\end{document}